\documentclass[aps,prd,onecolumn,groupedaddress,showpacs,nofootinbib,amssymb]{revtex4}

\usepackage{graphicx,bm}

\usepackage[english]{babel}

\usepackage{amsmath}

\usepackage{amssymb}

\usepackage{amsfonts}

\newcommand{\lp}{\left(}
\newcommand{\rp}{\right)}

\newcommand{\be}{\begin{equation}}

\newcommand{\ee}{\end{equation}}

\newcommand{\bea}{\begin{eqnarray}}

\newcommand{\eea}{\end{eqnarray}}

\newcommand{\beq}{\begin{eqnarray}}

\newcommand{\eeq}{\end{eqnarray}}

\newcommand{\R}{\tilde{R}}

\newcommand{\e}{\mbox{e}}

\newcommand{\rd}{\mathrm{d}}

\newcommand{\vect}[1]{\bm{#1}}

\newcommand{\trace}{\mathrm{Tr}}

\newcommand{\lambdaw}{\Lambda_W^{}}

\newcommand{\p}{\partial}

\begin{document}

\title{Ekpyrotic universes in $F(R)$ Ho\v{r}ava-Lifshitz gravity}

\author{Antonio L\'opez-Revelles$^{1}$, Ratbay Myrzakulov$^{2}$, Diego S\'aez-G\'omez$^{1,3}$} 
\affiliation{
$^1$ Institut de Ci\`encies de l'Espai (IEEC-CSIC),
Campus UAB, Facultat de Ci\`encies, Torre C5-Par-2a pl, E-08193 Bellaterra (Barcelona), Spain, EU \\
$^2$Dept. Gen. Theor. Phys., Eurasian National University, Astana, 010008, Kazakhstan, \\
$^3$ Fisika Teorikoaren eta Zientziaren Historia Saila, Zientzia eta Teknologia Fakultatea, Euskal Herriko Unibertsitatea, 644 Posta Kutxatila, 48080 Bilbao, Spain, EU}

\begin{abstract}

The Ekpyrotic scenario is studied in the context of some extensions of Ho\v{r}ava-Lifshitz gravity. Some particular solutions that lead to 
cyclic Hubble parameters are analyzed, where the corresponding gravitational actions are reconstructed by using several techniques and 
auxiliary fields. Its comparison with standard $F(R)$ gravity is performed. In addition, the so-called Little Rip, a stage of the universe 
evolution when some bounded systems may be dissolute, is also studied in this frame of theories. 
\end{abstract}

\maketitle

\section{Introduction}

During the last years, one of the main theoretical problems that has concerned to the scientific community refers to the mystery of the 
accelerating expansion of the universe. Since a deviation in the luminosity distance of Supernovae Ia was observed in 1999, along with other 
independent observations (such as Cosmic Microwave Background (CMB) anisotropies), the fact that the expansion is accelerating has been mostly accepted. Such acceleration
seems to be approximately an effect of a cosmological constant in the Einstein field equations, exactly the same form that the vacuum energy 
density acquires. Nevertheless, the observed value is so small in comparison with the one predicted by quantum field theories that the probable
existence  of a dark component in the universe has been established, which would be responsible for the accelerating expansion, and may deviate
in principle from a perfect de Sitter acceleration (in other words from a cosmological constant). Under the name of dark energy, plenty of 
candidates have been proposed which can perfectly predict the observational data, leading to a problem of degeneracy.  On the other hand, 
standard model of cosmology also requires an initial stage, the so-called inflation, in order to solve some cosmological problems as the 
homogeneity or the flatness problems. As an alternative, the so-called ekpyrotic scenario may avoid the need to provide initial conditions (inherent in every inflationary model), since the universe evolution acquires a periodic
behavior, such that in every cycle a new universe is born (see Ref.~\cite{ekpyrotic}). In addition, it is argued that the problem of flatness does not appear
in this model because the universe initially
was in a nearly BPS (BogolmonÕyi-Prasad-Sommerfield) state, which is homogeneous (see Ref.~\cite{ekpyrotic}). In the last years, very promising models capable to unify 
the entire cosmic evolution under the same mechanism have been proposed, where the  inflationary epoch  and the late-time acceleration era are 
unified under the same mechanism (or alternatively the ekpyrotic scenario), providing a simpler picture of the universe evolution. Most such 
models are described by scalar fields due to its simple form (see Ref.~\cite{Elizalde:2008yf} and references therein), or other kind of fields 
(see Ref.~\cite{Elizalde:2010xq}), but also a large effort has been done in the reconstruction of modified gravity theories (for a general report, see Ref.~\cite{Clifton:2011jh}) available to 
reproduce the cosmic evolution (for a review, see Ref.~\cite{review}, and Refs.~\cite{Nojiri:2006gh,SaezGomez:2008uj}), which may seem more 
natural as they are expressed in terms purely of the metric tensor without additional fields.\\

On the other hand, a new theory of gravity that is power-counting renormalizable has been proposed recently in Ref.~\cite{Horava}. Such theory, 
already known as Ho\v{r}ava-Lifshitz gravity, breaks the invariance under full diffeomorphisms of General Relativity by introducing an 
anisotropy between the spatial and time coordinates through a critical exponent $z$. This restriction of the symmetries allows the theory to be 
power-counting renormalizable, but an additional scalar degree of freedom is found, which introduces instabilities in the spectrum of the 
theory (see Refs.~\cite{Charmousis:2009tc,Blas:2009yd}). However, some extensions of the theory seem to address the problem of the scalar mode \cite{Blas:2009qj,Horava3}, as well as to 
generalize the action to more complex ones (see Ref.~\cite{Kluson:2010za}). Moreover,  cosmological models have been widely studied in the 
context of Ho\v{r}ava-Lifshitz gravity (see Ref.~\cite{cosm}), and also  generalizations of the original action (similarly to standard $F(R)$ 
gravity) have been proposed, where the entire cosmological history can be well reproduced, and it has also a good UV behavior (see Refs.\cite{FRhorava,Elizalde:2010ep}).  \\

The aim of the present paper is to study the ekpyrotic scenario in the frame of some extensions of Ho\v{r}ava-Lifshitz gravity, where a 
universe described in terms purely of gravity is able to pass along the different stages of an ekpyrotic model. This class of cosmological 
solutions can be realized in standard $F(R)$ gravity as shown in Ref.~\cite{Nojiri:2011kd}. Here, we reconstruct some periodic solutions for 
the Hubble parameter, which may be able to describe the entire evolution of the universe. In addition, we also analyze the shape of the action 
for each phase of the ekpyrotic scenario, where the possibility of the occurrence of a {\it Little Rip} is  explored. The so-called {\it Little 
Rip} is a postulated phase of the universe evolution, when a very strong accelerating expansion would lead  to break some  bounded systems, as 
the Solar System or even the molecules and atoms (see Ref.~\cite{LittleRip}). Such breaking is shown to be fully compatible with the ekpyrotic 
scenario in comparison with future singularities as the Big Rip that are not, unless some cure for the future singularity is considered 
\cite{LopezRevelles:2011uc}. Moreover, the presence of a Big Bang/Crunch singularity, usual in ekpyrotic cosmologies, is still an open issue for this kind of cyclic scenario, where quantum effects may resolve it (see \cite{Turok:2004gb}) or an effective theory that generates a non singular bounce (see \cite{Buchbinder:2007ad}). Nevertheless, here we are interested to explore the classical effects of the theory, where some non singular solutions are proposed, while the study of possible UV effects in the presence of the singularity is beyond the purpose of this paper.  \\

The paper is organized as follows: in the next section, $F(R)$ Ho\v{r}ava-Lifshitz gravity is briefly reviewed. In section III, the actions for 
some cyclic solutions are reconstructed. Finally, section IV is devoted to the analysis of ekpyrotic scenario, where each phase of the cycle is 
analyzed.

\section{Modified $F(R)$ Ho\v{r}ava-Lifshitz gravity \label{II}}

In this section, modified Ho\v{r}ava-Lifshitz $F(R)$
gravity is briefly reviewed \cite{Kluson:2010za,FRhorava,Elizalde:2010ep}. We start by writing a
general metric in the so-called Arnowitt-Deser-Misner (ADM) decomposition in a $3+1$ spacetime (for
more details see \cite{ADM}),
\be
ds^2=-N^2 dt^2+g^{(3)}_{ij}(dx^i+N^idt)(dx^j+N^jdt)\, ,
\label{1.1}
\ee
where $i,j=1,2,3$, $N$ is the so-called lapse variable, and $N^i$ is
the shift $3$-vector. In standard general relativity (GR),
the Ricci scalar can be written in terms of this metric, and yields
\be
R=K_{ij}K^{ij}-K^2+R^{(3)}+2\nabla_{\mu}(n^{\mu}\nabla_{\nu}n^{\nu}-n^{\nu}
\nabla_{\nu}n^{\mu})\, ,
\label{1.2}
\ee
here $K=g^{ij}K_{ij}$, $K_{ij}$ is the extrinsic curvature, $R^{(3)}$
is the spatial scalar curvature, and $n^{\mu}$ a unit vector
perpendicular to a hypersurface of constant time. The extrinsic
curvature $K_{ij}$ is defined as
\be
K_{ij}=\frac{1}{2N}\left(\dot{g}_{ij}^{(3)}-\nabla_i^{(3)}N_j-\nabla_j^{(3)}
N_i\right)\, .
\label{1.3}
\ee

In the original model \cite{Horava}, the lapse variable $N$ is taken
to be just time-dependent, so that the projectability condition holds
and by using the foliation-preserving diffeomorphisms (\ref{1.7}),
it can be fixed to be $N=1$.
As pointed out in \cite{Blas:2009qj},  imposing the projectability
condition may cause problems with Newton's law in the Ho\v{r}ava gravity.
For the non-projectable case, the Newton law
could be restored (while keeping stability) by the ``healthy''
extension of the original Ho\v{r}ava gravity of Ref.~\cite{Blas:2009qj}.

The action for standard $F(R)$ gravity can be written as
\be
S=\int d^4x\sqrt{g^{(3)}}N F(R)\, .
\label{1.4}
\ee
Gravity of Ref.~\cite{Horava} is assumed to have different scaling
properties of the space and time coordinates
\be
x^i=b x^i\, , \quad t=b^zt\, ,
\label{1.6}
\ee
where $z$ is a dynamical critical exponent that renders the theory
renormalizable for $z=3$ in $3+1$ spacetime dimensions \cite{Horava}.
GR is recovered when $z=1$. The scaling properties (\ref{1.6}) render
the theory  invariant only under the so-called foliation-preserving 
diffeomorphisms:
\be
\delta x^i=\zeta(x^i,t)\, , \quad \delta t=f(t)\, .
\label{1.7}
\ee
It has been pointed that, in the IR limit, the additional scalar degree of freedom
can be removed by means of an additional $U(1)$ symmetry \cite{Horava3}.  Here, we are interested on actions as follow,
\be
S=\frac{1}{2\kappa^2}\int dtd^3x\sqrt{g^{(3)}}N F(\tilde{R})\, , \quad
\tilde{R}= K_{ij}K^{ij}-\lambda K^2 + R^{(3)}+
2\mu\nabla_{\mu}(n^{\mu}\nabla_{\nu}n^{\nu}-n^{\nu}\nabla_{\nu}n^{\mu})-
L^{(3)}(g_{ij}^{(3)})\, ,
\label{1.8}
\ee
where $\kappa$ is the dimensionless gravitational coupling, and where, two
new constants $\lambda$ and $\mu$ appear, which account for the violation
of the full diffeomorphism transformations.
Note that in the original Ho\v{r}ava gravity theory \cite{Horava},
the fourth term in the expression for $\tilde{R}$ can be omitted, as
it becomes a total derivative. This generalization of the Ho\v{r}ava-Lifshitz action, similar to standard $F(R)$ gravity, may provide the way to describe the entire cosmological evolution with no need to introduce any additional field but where an additional scalar mode is assumed.  The possibility of violations of Newtonian law, due to the extra scalar mode coming from $F(\tilde{R})$, can be avoided by the appropriate expression for the action, as it was pointed out in Ref.~\cite{Elizalde:2010ep}. In addition, standard $F(R)$ gravity \eqref{1.4} can be recovered by setting $\lambda=\mu=1$. The term $L^{(3)}(g_{ij}^{(3)})$ in the action \eqref{1.8} is
chosen to be \cite{Horava}
\be
L^{(3)}(g_{ij}^{(3)})=E^{ij}G_{ijkl}E^{kl}\, ,
\label{1.9}
\ee
where the generalized De Witt metric is given by,
\be
G^{ijkl}=\frac{1}{2}\left(g^{(3)ik}g^{(3)jl}+g^{(3)il}g^{(3)jk}\right)-\lambda g^{(3)ij}g^{(3)kl}\, .
\label{1.10}
\ee
In Ref.~\cite{Horava}, the expression for $E_{ij}$ is constructed to
satisfy the ``detailed balance principle'' in order to restrict the
number of free parameters of the theory, and it is defined through the
variation of an action
\be
\sqrt{g^{(3)}}E^{ij}=\frac{\delta W[g_{kl}]}{\delta g_{ij}}\, ,
\label{1.11}
\ee
The action $W[g_{kl}]$ is assumed to be defined by the metric and the covariant derivatives on the three-dimensional hypersurface $\sum_t$.
In \cite{Horava}, $W[g^{(3)}_{kl}]$ is explicitly given for the
case $z=2$,
\be
\label{HLF7c} W=\frac{1}{\kappa_W^2}\int
\rd^3\vect{x}\,\sqrt{g^{(3)}}(R-2\lambdaw)\, ,
\ee
and for the case $z=3$,
\be
\label{HLF7d}
W=\frac{1}{w^2}\int_{\Sigma_t}\omega_3(\Gamma)\, .
\ee
Here $\kappa_W$
in (\ref{HLF7c}) is a coupling constant of dimension $-1/2$ and
$w^2$ in (\ref{HLF7d}) is the dimensionless coupling constant.
$\omega_3(\Gamma)$ in (\ref{HLF7d}) is given by
\be
\label{HLF7e}
\omega_3(\Gamma) = \trace\left(\Gamma\wedge
d\Gamma+\frac{2}{3}\Gamma\wedge\Gamma \wedge\Gamma\right) \equiv
\varepsilon^{ijk}\left(\Gamma^{m}_{il}\p_j
\Gamma^{l}_{km}+\frac{2}{3}\Gamma^{n}_{il}\Gamma^{l}_{jm}
\Gamma^{m}_{kn}\right)\rd^3\vect{x}\, .
\ee
Here we are interested in the study of cosmological
solutions for the theory described by action (\ref{1.8}).
Spatially-flat Friedmann-Lema\^itre-Robertson-Walker (FLRW) metric is assumed
\be
ds^2=-N^2dt^2+a^2(t)\sum_{i=1}^3 \left(dx^{i}\right)^2\, ,
\label{1.14}
\ee
where $N$ is taken to be just time-dependent (projectability condition) and, by using the
foliation-preserving
diffeomorphisms (\ref{1.7}), it can be set to  unity, $N=1$. Then, just as an assumption of
the solution, $N$ is taken to be unity.

For a flat FLRW metric (\ref{1.14}), and a vanishing cosmological constant, the scalar $\tilde{R}$ is
given by
\be
\tilde{R}=\frac{3(1-3\lambda
+6\mu)H^2}{N^2}+\frac{6\mu}{N}\frac{d}{dt}\left(\frac{H}{N}\right)\, .
\label{1.15}
\ee
For the action (\ref{1.8}), and assuming the FLRW metric (\ref{1.15}),
the second FLRW equation can be obtained by varying
the action with respect to the spatial metric $g_{ij}^{(3)}$, what
yields
\be
0=F(\tilde{R})-2(1-3\lambda+3\mu)\left(\dot{H}+3H^2\right)F'(\tilde{R})-2(1-
3\lambda)H
\dot{\tilde{R}}F''(\tilde{R})+2\mu\left(\dot{\tilde{R}}^2F^{(3)}(\tilde{R})
+\ddot{\tilde{R}}F''(\tilde{R})\right)+\kappa^2p_m\, ,
\label{1.16}
\ee
here $\kappa^2=16\pi G$, $p_m$ is the pressure of a perfect fluid
that fills the universe, and $N=1$. Note that this
equation turns out the usual second FLRW equation for standard $F(R)$
gravity (\ref{1.4}) when  $\lambda=\mu=1$.
If we assume the projectability condition,
variation over $N$ of the action (\ref{1.8}) yields the following
global constraint
\be
0=\int d^3x\left[F(\tilde{R})-6F'(\tilde{R})\left\lbrace(1-3\lambda +3\mu)H^2+\mu\dot{H}\right\rbrace +6\mu
H \dot{\tilde{R}}F''(\tilde{R})-\kappa^2\rho_m\right]\, .
\label{1.17}
\ee
Now,  by using the ordinary conservation equation for the matter fluid
$\dot{\rho}_m+3H(\rho_m+p_m)=0$, and integrating Eq.~(\ref{1.16}), it yields
\be
0=F(\tilde{R})-6\left[(1-3\lambda
+3\mu)H^2+\mu\dot{H}\right]F'(\tilde{R})+6\mu H
\dot{\tilde{R}}F''(\tilde{R})-\kappa^2\rho_m-\frac{C}{a^3}\, ,
\label{1.18}
\ee
where $C$ is an integration constant, taken to be zero,
according to the constraint equation (\ref{1.17}). In
\cite{Mukohyama:2009mz}, however, it has been claimed that $C$ needs
not always vanish in a local region, since (\ref{1.17}) needs to be
satisfied in the whole universe. In the region $C>0$, the $Ca^{-3}$
term in (\ref{1.18}) may be regarded as dark matter.

If we do not assume the projectability condition, we can directly
obtain (\ref{1.18}), which corresponds to the first FLRW equation, by 
varying the action (\ref{1.8}) over $N$.
Hence, starting from a given $F(\tilde{R})$ function, and solving 
Eqs.~(\ref{1.16}) and (\ref{1.17}), a cosmological solution
can be obtained.

\section{Reconstructing  cyclic universes}

    The aim of this section is to show that any cosmology may be realized in $F (\tilde R)$ Ho\v{r}ava-Lifshitz gravity. For this purpose, we 
    present two different methods of reconstruction, the first one is based on the use of the number of e-foldings,  while, the second one uses an auxiliary scalar field.

    \subsection{Reconstructing  cyclic universe using e-folding}

    We will assume the flat FLRW 
    metric defined in \eqref{1.14} with $N=1$, in such a case the first FLRW equation is given by \eqref{1.18} with $C=0$, which can be 
    rewritten as a function of the number of e-foldings $\eta=\ln\frac{a}{a_0}$ instead of the time $t$. This technique has been developed in 
    \cite{Nojiri:2009kx} for classical $F(R)$ gravity, and for Ho\v{r}ava-Lifshitz F(R)-gravity \cite{Elizalde:2010ep}. Since 
    $\frac{d}{dt}=H\frac{d}{d\eta}$ and $\frac{d^2}{d\eta^2}=H^2\frac{d^2}{d\eta^2}+H\frac{dH}{d\eta}\frac{d}{d\eta}$, the first FLRW equation 
    (22) is rewritten as
      \be
	0=F(\tilde{R})-6\left[\frac{A}{3}H^2+\mu HH'\right]\frac{dF(\tilde{R})}{d\tilde{R}}+6\mu H^2\left[2AHH'+6\mu H'^2+6\mu H''H' \right] 
	\frac{d^2F(\tilde{R})}{d^2\tilde{R}}-\rho\ ,
	\label{D1}
      \ee   
    where $A=3-9\lambda+18\mu$ and the primes denote derivatives respect $\eta$. By using the energy conservation equation 
    $\dot{\rho}+3H(1+w)\rho=0$, the energy density yields,
      \be
	\rho=\rho_0a^{-3(1+w)}=\rho_0a_0^{-3(1+w)}e^{-3(1+w)\eta}\ .
	\label{D2}
      \ee
    As the Hubble parameter can be written as a function of the number of e-foldings, $H=H(\eta)$, the  scalar curvature in \eqref{1.15} takes 
    the form
      \be
	\tilde{R}=AH^2+6\mu HH'\ ,
	\label{D3}
      \ee
    which can be solved respect to $\eta$ as $\eta=\eta(\tilde{R})$. Then, the equation (\ref{D1}) for $F(\tilde{R})$ with the variable 
    $\tilde{R}$ is obtained. This can be a little simplified by writing $G(\eta)=H^2$ instead of the Hubble parameter. In such a case, the 
    differential equation (\ref{D1}) gives
      \be
	0=F(\tilde{R})-6\left[\frac{A}{3}G+\frac{\mu}{2}G'\right]\frac{dF(\tilde{R})}{d\tilde{R}}+6\mu\left[AGG'+3\mu GG''\right] \frac{d^2F(\tilde{R})}{d^2\tilde{R}}-\rho_0a_0^{-3(1+w)}e^{-3(1+w)\eta}\ ,
	\label{D4}
      \ee
    and the scalar curvature is now written as $\tilde R = A G + 3 \mu G'$. Hence, for a given cosmological solution $H^2=G(\eta)$, one can solve the 
    equation (\ref{D4}), and the corresponding $F(\tilde{R})$ is obtained.\\
    
    In order to illustrate that cyclic solutions can be reproduced by this kind of theories, let us consider the following example:
      \begin{equation}\label{1}
	H(t) = - \frac{2 \pi}{T} H_1 \sin{\lp \frac{2 \pi}{T} t \rp}
      \end{equation}
    where  $H_1$ and $T$ are constants. The number of e-foldings is:
                \begin{equation}\label{4}
                    H(t) = \frac{1}{a} \frac{da}{dt} = - \frac{2 \pi}{T} H_1 \sin{\lp \frac{2 \pi}{T} t \rp}
                    \Longrightarrow \frac{da}{a} = - \frac{2 \pi}{T} H_1 \sin{\lp \frac{2 \pi}{T} t \rp} dt
                    \Longrightarrow$$
                    $$\Longrightarrow \eta (t) = \ln{\lp \frac{a(t)}{a_0} \rp} = H_1 \left[ \cos{\lp \frac{2 \pi}{T}
                    t \rp} - 1 \right]
                \end{equation}
            Using (\ref{4}), the function $G(\eta)$ and its derivatives are given by:
                \begin{equation}\label{5}
                        G(\eta) = H^2 = - \lp \frac{2 \pi}{T} \rp^2 \lp 2 H_1 + \eta \rp \eta, \quad
                        G'(\eta) = - 2 \lp \frac{2 \pi}{T} \rp^2 \lp H_1 + \eta \rp, \quad
                                               G''(\eta) = - 2 \lp \frac{2 \pi}{T} \rp^2.
                    \end{equation}
            Then, we have:
                \begin{equation}\label{6}
                    \tilde R = - 3 \lp \frac{2 \pi}{T} \rp^2 \left[ 2 \mu H_1 + 2 \eta \lp \mu + \lp 1 - 3 \lambda + 6 \mu
                    \rp H_1 \rp + \lp 1 - 3 \lambda + 6 \mu \rp \eta^2
                    \right] \Rightarrow$$
                    $$\Rightarrow \eta = - \lp \frac{\mu}{1 - 3 \lambda + 6 \mu} + H_1
                    \rp \pm \sqrt{\frac{\mu^2}{(1 - 3 \lambda + 6 \mu)^2} + H_1^2 - \frac{\tilde R}{3 (1 - 3 \lambda + 6
                    \mu)\lp \frac{2 \pi}{T} \rp^2}}.
                \end{equation}
            Now, if we call $x = \pm \sqrt{\frac{\mu^2}{(1 - 3 \lambda + 6 \mu)^2} + H_1^2 - \frac{\tilde R}{3 (1 - 3
            \lambda + 6 \mu) \lp \frac{2 \pi}{T} \rp^2}}$, we can write:
                \begin{equation}\label{7}
                    \eta = - \lp \frac{\mu}{1 - 3 \lambda + 6 \mu} + H_1 \rp + x.
                \end{equation}
            We also have that:
                \begin{equation}\label{8}
                    \frac{dF(\tilde R)}{d \tilde R} = - \frac{1}{6 (1 - 3 \lambda + 6 \mu) \lp \frac{2 \pi}{T}
                    \rp^2 x} \frac{dF_1 (x)}{dx},$$
                    $$\frac{d^2 F(\tilde R)}{d \tilde R^2} = \frac{1}{\left[ 6 (1 - 3 \lambda + 6 \mu) \lp
                    \frac{2 \pi}{T} \rp^2 x \right]^2} \lp \frac{d^2 F_1 (x)}{dx^2} - \frac{1}{x}
                    \frac{dF_1 (x)}{dx} \rp,
                \end{equation}
            where $F_1 (x) = F(\tilde R(x))$.

            We can now rewrite (\ref{5}) in terms of the new variable $x$ by using (\ref{7}), leading to:
                \[ 
                        G(\eta(x)) = - \lp \frac{2 \pi}{T} \rp^2 \lp \frac{\mu^2}{(1 - 3 \lambda + 6 \mu)^2}
                        - H_1^2 - \frac{2 \mu}{1 - 3 \lambda + 6 \mu} x + x^2 \rp, \]
                        \[
                        G'(\eta(x)) = - 2 \lp \frac{2 \pi}{T} \rp^2 \lp - \frac{\mu}{1 - 3 \lambda + 6 \mu}
                        + x \rp, \]
                        \be
                        G''(\eta(x)) = - 2 \lp \frac{2 \pi}{T} \rp^2.
                                    \label{9}
                \ee
            Finally, by introducing (\ref{8}-\ref{9}) into the equation (\ref{D4}) and considering the case of vacuum, we arrive to the following differential equation for $F_1 (x)$:
                \begin{equation}\label{11}
                    0 = x^2 F_1 (x) + \left[ \frac{\mu}{1 - 3 \lambda + 6 \mu} \lp H_1^2 - \frac{\mu^2}{(1 - 3
                    \lambda + 6 \mu)^2} \rp + \lp \frac{2 \mu^2}{(1 - 3 \lambda + 6 \mu)^2} + H_1 \rp x - x^3
                    \right] \frac{dF_1 (x)}{dx} +$$
                    $$+ \frac{\mu}{1 - 3 \lambda + 6 \mu} x \left[ - \lp H_1^2 -
                    \frac{\mu^2}{(1 - 3 \lambda + 6 \mu)^2} \rp - \frac{2 \mu}{1 - 3 \lambda + 6 \mu} x + x^2
                    \right] \frac{d^2 F_1 (x)}{dx^2}\ .
                \end{equation}
Here, we have obtained an equation for the gravitational action, that in principle can not provide an exact expression, but which can be integrated numerically. Hence, this solution reproduces a periodic behavior for the Hubble parameter leading to a cyclic  universe. 
   
    \subsection{Reconstructing cyclic universe using a scalar field}

  In this subsection it will be shown how to construct an $F (\tilde R)$ Ho\v{r}ava-Lifshitz gravity model realizing any given cosmology, this time using instead the technique of 
  \cite{Nojiri:2006gh}. We start from the action for $F (\tilde R)$ Ho\v{r}ava-Lifshitz gravity
    \begin{equation}\label{t12}
      S = \int dt d^3 x \sqrt{g^{(3)}} N (F(\tilde R) + \mathcal{L}_{matter}),
    \end{equation}
  which is equivalent to 
    \begin{equation}\label{t13}
      S = \int dt d^3 x \sqrt{g^{(3)}} N (P(\phi) \tilde R + Q(\phi) + \mathcal{L}_{matter}).
    \end{equation}
  Here, $\mathcal{L}_{matter}$ is the matter Lagrangian density and $P$ and $Q$ are proper functions of the scalar field, $\phi$, which can be 
  regarded as an auxiliary field, because there is no kinetic term depending on $\phi$ in the Lagrangian. By varying the action with respect 
  to $\phi$, it follows that
    \begin{equation}\label{t14}
      0 = P'(\phi) \tilde R + Q'(\phi),
    \end{equation}
  which can be solved in terms of $\phi$, as
    \begin{equation}\label{t15}
      \phi = \phi (\tilde R).
    \end{equation}
  By substituting (\ref{t15}) into (\ref{t13}) and comparing with (\ref{t12}), one obtains
    \begin{equation}\label{t16}
      S = \int dt d^3 x \sqrt{g^{(3)}} N (F( \tilde R) + \mathcal{L}_{matter}),$$
      $$ F(\tilde R) \equiv P(\phi(\tilde R)) \tilde R + Q(\phi(\tilde R)).
    \end{equation}
  We proceed now in the same way that we did in Section II, assuming the FLRWL metric, the second FLRWL equation can be obtained by varying the 
  action (\ref{t13}) with respect to the spatial metric $g^{(3)}_{ij}$. This equation can be written as:
    \begin{equation}\label{t17}
      P(\phi) \left\{ \tilde R - 2 \left( 1 - 3 \lambda + 3 \mu \right) \left( 3 H^2 + \dot H \right) \right\} - 2 \left( 1 - 3 \lambda \right)
      H \frac{d P(\phi)}{dt} + 2 \mu \frac{d^2 P(\phi)}{dt^2} + Q(\phi) + p = 0
    \end{equation}
  If we assume now the projectability condition, we can obtain a global constraint doing the variation of the action (\ref{1.8}) over $N$, it 
  yields:
    \begin{equation}\label{t18}
      P(\phi) \left\{ \tilde R - 6 \left[ \left( 1 - 3 \lambda + 3 \mu \right) H^2 + \mu \dot H \right] \right\} + 6 \mu H 
      \frac{d P(\phi)}{dt} + Q(\phi) - \rho = 0
    \end{equation}
  We can combine (\ref{t17}) and (\ref{t18}) in order to eliminate the function $Q(\phi)$, we finally obtain:
    \begin{equation}\label{t20}
      2 \mu \frac{d^2 P(\phi(t))}{dt^2} - 2 \left( 1 - 3 \lambda + 3 \mu \right) H \frac{d P(\phi(t))}{dt} - 2 \left( 1 - 3 \lambda \right) 
      \dot H  P(\phi(t)) + p + \rho = 0
    \end{equation}
  As we may redefine the scalar field $\phi$ properly, we can choose
    \begin{equation}\label{t21}
      \phi = t.
    \end{equation}
  Provided the scale factor $a$ is given by a proper function $g(t)$ as
    \begin{equation}\label{t22}
      a = a_0 e^{g(t)},
    \end{equation}
  with a constant $a_0$, and if it is moreover assumed that $p$ and $\rho$ are the sum of the different matter contributions, with constant equation of state (EoS) 
  parameters $\omega_i$, Eq.~(\ref{t20}) then reduces to the following second order differential equation
    \begin{equation}\label{t23}
      2 \mu \frac{d^2 P(\phi)}{d \phi^2} - 2 \left( 1 - 3 \lambda + 3 \mu \right) g'(\phi) \frac{d P(\phi)}{d \phi} - 2 \left( 1 - 3 \lambda 
      \right) g''(\phi)  P(\phi) + \sum \limits_i (1 + \omega_i) \rho_{i0} a_0^{-3 (1 + \omega_i)} e^{-3 (1 + \omega_i) g(\phi)} = 0
    \end{equation}
  From this equation we can obtain $P(\phi)$ and using Eq.~(\ref{t18}) we find that
    \begin{equation}\label{t24}
      Q(\phi) = - P(\phi) \left\{ \tilde R - 6 \left[ \left( 1 - 3 \lambda + 3 \mu \right) H^2 + \mu \dot H \right] \right\} - 6 \mu H 
      \frac{d P(\phi)}{dt} + \sum \limits_i \rho_{i0} a_0^{-3 (1 + \omega_i)} e^{-3 (1 + \omega_i) g(\phi)}
    \end{equation}
  As a result, any given cosmology, expressed as (\ref{t22}), can indeed be realized (as anticipated) by some specific $f(R)$-gravity. Note 
  that Eq.(\ref{t23}) is a second order differential equation on $P(\phi)$ when $g'(\phi)$ is known, but it can also be considered as a first 
  order differential equation on $g'(\phi)$ (i.e. on $H(\phi)$) in the case that the function $P(\phi)$ is given. In the following we will use
  this last point of view to find out a function $F(\tilde R)$ that reproduces a  cyclic universe.
  
  When matter can be neglected Eq.(\ref{t23}) can be rewritten as:
    \begin{equation}\label{t25}
      \frac{d}{d \phi} \left( g'(\phi) \, P(\phi)^{\frac{1 - 3 \lambda + 3 \mu}{1 - 3 \lambda}} \right) = \frac{\mu}{1 - 3 \lambda} \,
      P(\phi)^{\frac{3 \mu}{1 - 3 \lambda}} \, \frac{d^2 P(\phi)}{d \phi^2}
    \end{equation}
  which can be solved as \cite{SaezGomez:2008uj}:
    \begin{equation}\label{t26}
      g'(\phi) = \frac{\mu}{1 - 3 \lambda} \, P(\phi)^{- \frac{1 - 3 \lambda + 3 \mu}{1 - 3 \lambda}} \int d \phi \, 
      P(\phi)^{\frac{3 \mu}{1 - 3 \lambda}} \, \frac{d^2 P(\phi)}{d \phi^2} =$$
      $$= \frac{\mu}{1 - 3 \lambda} \, \frac{1}{P(\phi)} \, 
      \frac{dP(\phi)}{d \phi} - \frac{3 \mu^2}{( 1 - 3 \lambda )^2} \, P(\phi)^{- \frac{1 - 3 \lambda + 3 \mu}{1 - 3 \lambda}} \int d \phi \, 
      P(\phi)^{\frac{3 \mu}{1 - 3 \lambda} - 1} \, \left( \frac{dP(\phi)}{d \phi} \right)^2
    \end{equation}
  In the second equality, we have used the partial integration. Furthermore by writing $P(\phi)$ as:
    \begin{equation}\label{t27}
      P(\phi) = U(\phi)^{\frac{2 ( 1 - 3 \lambda )}{1 - 3 \lambda + 3 \mu}}
    \end{equation}
  (\ref{t26}) is rewritten as follows:
    \begin{equation}\label{t28}
      g'(\phi) = \frac{2\mu}{1 - 3 \lambda + 3 \mu} \, \frac{1}{U(\phi)} \, 
      \frac{dU(\phi)}{d \phi} - \frac{12 \mu^2}{( 1 - 3 \lambda + 3 \mu )^2} \, \frac{1}{U(\phi)^2} \int d \phi \, \left( 
      \frac{dU(\phi)}{d \phi} \right)^2.
    \end{equation}

  We now consider the case given by:
    \begin{equation}\label{t29}
      P(\phi) = U(\phi)^{\frac{2 ( 1 - 3 \lambda )}{1 - 3 \lambda + 3 \mu}} = 
      P_0 \left[ \cos(\omega \phi) \right]^{- \frac{2 ( 1 - 3 \lambda )}{1 - 3 \lambda + 3 \mu}}
    \end{equation}
  where $P_0$ and $\omega$ are constants. Then, using Eq.(\ref{t25}), the solution is given by:
    \begin{equation}\label{t30}
      g'(\phi) = g_0 \left[ \cos(\omega \phi) \right]^2 + \frac{2 \omega \mu}{1 - 3 \lambda + 3 \mu} \tan(\omega \phi) \left( 1 - 
      \frac{2 \mu}{1 - 3 \lambda + 3 \mu} \left[ \sin(\omega \phi) \right]^2 \right)
    \end{equation}
  where $g_0$ is an integration constant. Note that the tangent term in (\ref{t30}) makes the solutions to contain some divergences that 
  correspond to points where the scale factor becomes null, i.e. $a(t_0) = 0$. These divergences can be identified with a Big Bang/Crunch 
  singularity and they are very common in cyclic universes, where the ekpyrotic scenario is reproduced. In order to have a smooth transition through the Big Bang/Crunch singularity, one expects that the quantum effects of the theory will avoid the occurrence of the singularity. However, this is a large task, even more in a background solution as \eqref{t30}, and should be explored separately in the future. In addition, other mechanisms for a smooth transition have been suggested as the introduction of an additional term in the action or a different coupling with the matter lagrangian (see Ref.~\cite{Khoury:2004xi}).

\section{Ekpyrotic scenario in Ho\v{r}ava-Lifshitz gravity}

We have shown above that periodic solutions can be easily reconstructed in the frame of extended Ho\v{r}ava-Lifshitz gravity. Here we are more interested to analyze ekpyrotic models in such kind of theories. The so-called Ekpyrotic/cyclic
universe is an alternative explanation to the inflationary paradigm proposed one decade ago in  Ref.~\cite{ekpyrotic}, that can provide a realistic picture of the universe evolution (for a confrontation between both models, see \cite{Linde:2002ws}). In the same way as the inflationary scenario, ekpyrotic cosmological models can also predict the origin of primordial inhomogeineties  that leads to the formation of large structures and the anisotropies observed in the CMB. In addition, this model does not require initial conditions in comparison with the standard inflationary scenario due to its cyclic nature.
In general, the cosmological evolution presented
by an ekpyrotic universe consist of infinite cycles, where each cycle contains four stages: a
first initial hot state
similar to the standard Big Bang model, then a phase of accelerated expansion,
after which the universe starts to contract  and finally the cycle ends in a Big
Bang/Crunch transition,
when the cycle starts again. The cosmological problems  enumerated above are solved during the contracting phase. In the usual ekpyrotic models, brane scenarios or
scalar fields are
considered (see \cite{ekpyrotic}). However, it is clear that modified
gravity,
and precisely $F(\R)$ gravity, can perfectly reproduce the ekpyrotic scenario \cite{Nojiri:2011kd}. Here we are  interested to see how the cosmological problems can be solved during the contracting phase in the context of Hor\v{r}ava-Lifshitz gravity, and to reconstruct the corresponding behavior of the action during each phase of an ekpyrotic universe. The first FLRW equation is given by,
\be
\frac{3}{\kappa^2}H^2=\frac{1}{(1-3\lambda+3\mu)F'(\R)}\left(\frac{\rho_{m0}}{a^3}+\frac{\rho_{r0}}{a^4}+\frac{\rho_{\sigma0}}{a^6}-\frac{k}{a^2}\right)+\rho_{F(\R)}\ ,
\label{E1}
\ee
where the subscripts refers to matter (m), radiation (r), anisotropies ($\sigma$), and $k$ is the spatial curvature, while $\rho_{F(R)}$ is defined as,
\be
\rho_{F(\R)}= \frac{1}{\kappa^2(1-3\lambda+3\mu)F'(\R)}\left(\frac{1}{2}F(\R)-3\mu\dot{H}F'(\R)+3\mu H\dot{\R}F''(\R)\right)\ .
\label{E2}
\ee
In order to solve the initial cosmological problems, the last term in (\ref{E1}) should dominate over the rest when the scale factor tends to zero, i.e. when the universe approaches the Big Bang (Crunch) singularity. Hence, the effective energy density defined in \eqref{E2} should behave as $\rho_{F(\R)}\propto1/a^m$ with $m>6$ when the scale factor tends to zero, such that close to the initial singularity, the FLRW equation \eqref{E1} can be approximated as,
\be
\frac{3}{\kappa^2}H^2\sim\rho_{F(\R)}\sim\frac{C}{a^m}\ ,
\label{E3}
\ee
where $C$ is a constant. Then, we can reconstruct the form of the action $F(R)$ close to the Big Bang (Crunch) singularity by solving the FLRW equation. Hence, for the Hubble parameter \eqref{E3}, the scalar curvature is given by,
\be
\R=\left[(1-3\lambda+6\mu)-\mu m\right]\kappa^2\frac{C}{a_m}\ .
\label{E4}
\ee
And the FLRW equation (\ref{E3}) yields an expression where $F(R)$ is the unknown quantity,
\be
\R^2F''(\R)+\frac{2\kappa^2(1-3\lambda+3\mu)-\mu m}{2\mu m}\R F'(\R)-\frac{(1-3\lambda+6\mu)-\mu m}{2\mu m}F(\R)=0\ .
\label{E5}
\ee
This is an Euler equation that can be easily solved, and gives the function for $F(R)$,
\be
F(\R)=\kappa_1\R^{\beta_+}+\kappa_2\R^{\beta_-}\ .
\label{E6}
\ee
where,
\be
\beta_{\pm}=\frac{3m\mu-2\kappa^2(1-3\lambda+3\mu)\pm\sqrt{4\kappa^2 (1-3\lambda+3\mu)(\kappa^2(1-3\lambda+3\mu)-3m\mu)+m\mu(8-24\lambda+(48+m)\mu)}}{4m\mu}\ .
\label{E7}
\ee
Note that the scalar curvature tends to infinity when $a\rightarrow 0$, and in such strong gravity regime, the parameters $\lambda$ and $\mu$ should be different than one, the limit of General Relativity, as the breaking of Lorentz invariance will be present in such kind of regimes, while it is recovered for the weak field systems. Moreover, in order to get a smooth transition along the singularity, the first derivative of $F(\R)$ should tend to infinity to ensure that the matter energy densities remain finite in \eqref{E1}, which can be easily achieved when $(\beta_{\pm}-1)<0$ in \eqref{E5}.  \\
After this contracting phase, the ekpyrotic model suggests that a hot initial state, similar to the Big Bang model, is created (in the original ekpyrotic model by the collision between  branes), and which may be  created by the decaying of the extra scalar modes coming from $F(R)$ in this class of theories. Nevertheless, this is beyond the purpose of this paper, where our aim is to show the approximated form that the action should look like for each phase of the cycle. Then, during the matter/radiation dominated epochs, the action may seem as the standard Hilbert-Einstein action with $F(\R)\sim\R$ and $\R=R$, i. e. the parameters responsible of the breaking of full diffeomorphisms should recover the values of GR, $\lambda=\mu\sim1$. The last phase for each cycle refers to an accelerating era, which may be described by the usual $\Lambda$CDM model, whose Hubble parameter can be written in terms of the number of e-foldings as,
\be
H^2 = H_0^2 + \frac{\kappa^2}{3}\rho_0 a^{-3} = H_0^2 + \frac{\kappa^2}{3}\rho_0 a_0^{-3} \mbox{e}^{-3\eta} \ .
\label{E8}
\ee 
where $H_0$ and $\rho_0$ are constants. In the frame of General Relativity, the terms in the r.h.s of equation (\ref{E8}) correspond to an effective cosmological constant $\Lambda=3H_0^2$ and to  a pressureless fluid. The corresponding $F(\tilde{R})$ can be reconstructed by following the steps described above. For this case the function $G(\eta)$ is given by
\be
G(\eta)=H_0^2 + \frac{\kappa^2}{3}\rho_0 a_0^{-3} \e^{-3\eta}\ . 
\label{E9}
\ee
And by using the expression for the scalar curvature $R=AG+3\mu G'$, the relation between $\tilde{R}$ and $\eta$ is obtained,
\be
\e^{-3\eta}=\frac{R-AH_0^2}{k(3+9(\mu-\lambda))}\ ,
\label{E10}
\ee
where $k=\frac{\kappa^2}{3}\rho_0 a_0^{-3}$. Then, by substituting (\ref{E9}) and (\ref{E10}) in the equation (\ref{D4}), one gets the following differential expression,
\be
\frac{1+3(\mu-\lambda)}{6\mu(1-3\lambda)}F(\tilde{R})-\left[\frac{1+3(\mu-\lambda)}{3\mu(1-3\lambda)}\tilde{R}-\frac{3H^2_0\mu(1-3\lambda+6\mu)}{2\mu(1-3\lambda)}\right]F'(\tilde{R})-(\tilde{R}-9\mu H^2_0)(\tilde{R}-3H^2_0(1-3\lambda+6\mu))F''(\tilde{R})=0\ ,
\label{E11}
\ee 
here  we have neglected the contribution of matter for simplicity. By performing a change of variable $x=\frac{R-9\mu H^2_0}{3H^2_0(1+3(\mu-\lambda))}$, the equation (\ref{E11}) can be easily identified as an hypergeometric differential equation,
\be
0=x(1-x)\frac{d^2 F}{dx^2} + \left(\gamma - \left(\alpha + \beta + 1\right)x\right)\frac{dF}{dx} - \alpha \beta F\ ,
\label{E12}
\ee
with the set of parameters $(\alpha,\beta,\gamma)$  given by
\be
\gamma=-\frac{1}{2(1+3(\mu-\lambda))}\ , \quad \alpha+\beta= \frac{1+\lambda(9\mu-1)}{3\mu(1-3\lambda)}\ , \quad \alpha\beta=-\frac{1+3(\mu-\lambda)}{6\mu(1-3\lambda)}\ .
\label{E13}
\ee
The solution of the equation (\ref{E12}) is a Gauss' hypergeometric function \cite{Elizalde:2010ep},
\be
F(x) = C_1 F(\alpha,\beta,\gamma;x) + C_2 x^{1-\gamma} F(\alpha - \gamma + 1, \beta - \gamma + 1,2-\gamma;x)\ .
\label{E14}
\ee  
where $C_1$ and $C_2$ are constants. Then,  this action reproduces the $\Lambda$CDM model described by the Hubble parameter (\ref{E8}) without including a cosmological constants. Note that this is the same result obtained in \cite{Nojiri:2009kx} for classical $F(R)$ gravity, although in this case the solution depends on the parameters of the theory $(\mu,\lambda)$ whose values differ from the classical theory. \\
Other kind of accelerating expansions can be also reconstructed in the context of this class of theories as showed in Ref.~\cite{Elizalde:2010ep}. However, due to the periodic behavior of ekpyrotic universes, models containing future singularities (usually phantom models) are not allowed in this kind of models unless a mechanism for avoiding the singularity is introduced. Nevertheless, a new class of phantom models that do not contain Big Rip singularities but only affects to  bound systems without reaching a singular point, the so-called Little Rip, has been proposed in Ref.~\cite{LittleRip}, and extended to modified gravities in Ref.~\cite{Nojiri:2011kd}. Basically, these cosmological models consist on a phantom-like evolution, free of future singularities but whose strong expansion breaks the bond of some coupling systems (as galaxies, solar systems, or even atoms, nuclei...), what has been called as a Little Rip.  An simple example of this kind of evolution can be described by the Hubble parameter,
\be
H(t)\sim H_0 t\ ,
\label{E15}
 \ee
where $H_0$ is a constant. In this case, we can also reconstruct the corresponding $F(\R)$ action by solving the FLRW equation \eqref{1.18}. The scalar curvature is given by,
\be
\R= 3 (1-3\lambda+6\mu) H_0^2t^2+6\mu H_0\ .
\label{E16}
\ee
Then, the FLRW \eqref{1.18} yields,
\be
\frac{1}{2}F(\R)-\left(3H_0\mu+\frac{(1-3\lambda+3\mu)(R-6H_0\mu)}{1-3\lambda+6\mu}\right)F'(\R)+6H_0\mu(R-6H_0\mu)F''(\R)=0\ .
\label{E17}
 \ee
 This is also an hypergeometric equation, whose solution is given by,
 \be
 F(\R)=\left[C_1U(\gamma,\beta;x(\R))+C_2 L_{\gamma}^{(\alpha)}(x(\R))\right](\R-6H_0\mu)^{3/2}\ ,
 \label{E18}
 \ee
 where $U(\gamma,\beta;x)$ is the confluent hypergeometric function and $L_{\gamma}^{(\alpha)}(x)$ is the Laguerre polynomial. The variable $x(\R)$ and the set of parameters $(\gamma, \beta, \alpha)$ are defined as,
 \be
 x(\R)=\frac{(1-3\lambda+3\mu)(R-6H_0\mu)}{6H_0\mu(1-3\lambda+6\mu)}\ , \quad \gamma=-\frac{(2-6\lambda+3\mu)}{2(1-3\lambda+3\mu)}\ , \quad \beta=\frac{5}{2}\ , \quad \alpha=\frac{3}{2}\ .
 \label{E19}
 \ee
 Hence, the $F(\R)$ action \eqref{E18} corresponds to a series of powers in $\R$ that are capable to reproduce a kind of behavior given by the Hubble parameter \eqref{E15}. In such  case, we have that the effective energy density can be approximated as, 
 \be
 \rho_{F(\R)}\propto t^2\ .
 \label{E20}
 \ee
Note that for a cyclic universe, as the ones studied in section above, the phase when the universe expansion is accelerated can be approximated by  \eqref{E15}, such that a Little Rip may occur in the ekpyrotic scenario. In order to show in a qualitative way how this Little Rip occurs, i.e. how some  bounded systems are broken, let us  compare the effective energy density \eqref{E20} with the energy density of some known systems as the Solar-Earth system, and calculate the time remaining before the Little Rip occurs. By assuming that $\rho_{F(R)}(t_0)=\frac{3}{\kappa^2}H_0^2\sim 10^{-47}\, \mathrm{GeV}^4$, where the age of the universe is taken to be $t_0\sim 13.73\, \mathrm{Gyrs}$, according to
Ref.~\cite{Spergel:2006hy}, and a mean density of the Sun-Earth system given by $\rho_{\odot-\oplus}=0.594\times 10^{-3}\, \mathrm{kg/m}^3 \sim 10^{-21}\, \mathrm{GeV}^4$,
according to the evolution \eqref{E20}, the time for the little rip  is, 
 \be
 t_{LR}\sim 10^{13} Gyrs\ ,
 \label{E21}
 \ee
which is a large period compared with the current age of the universe. For other kind of expansions, as the an exponential Hubble parameter (studied in \cite{Nojiri:2011kd}), this time can be much shorter ($\sim 300Gyrs$). However, in an ekpyrotic scenario the occurrence of a Little Rip will depend on the duration of the accelerating phase before this ends, and a new contracting phase starts again. Note also that close to the dissolution of the bound structure,  gravity will be very strong, and the breaking of Lorentz invariance will be present, such that the values of $(\lambda, \mu)$ will determine the expansion rate, and for instance the occurrence of the Little Rip. \\

Let us now  consider a model that may reproduce a entire cycle of an ekpyrotic universe, 
\begin{equation}
H=H_0-H_1e^{-\beta t}.
\label{M1}
\end{equation}
For $H_1>H_0$, the Hubble parameter \eqref{M1} represents a universe that crosses through out a contracting phase, and then ends in an 
accelerating expansion for large times. Obviously, one would need to provide the way to start a cycle again, however for a qualitative 
description, we assume here that the cycle starts again after the accelerating phase somehow. For the solution \eqref{M1}, we have
\begin{equation}
\tilde{R}=AH^2+6\mu\dot{H}=A(H_0^2-2H_0H_1e^{-\beta t}+H_1^2e^{-2\beta t})-6\mu \beta H_1e^{-\beta t}
\label{M2}
\end{equation}
where we recall that $A=3(1-3\lambda+6\mu)$. From  \eqref{M2} we get
\begin{equation}
e^{-\beta t}=\frac{(AH_0+3\mu\beta)\pm\sqrt{(AH_0+3\mu\beta)^2-(AH_0^2-\tilde{R})}}{H_1}
\label{M3}
\end{equation}
For simplicity we consider the case when $AH_0+3\mu\beta=0$. Then Eq.~\eqref{M3} gives
\begin{equation}
e^{-\beta t}=\pm\frac{\sqrt{\tilde{R}-AH_0^2}}{H_1}\ .
\label{M4}
\end{equation}
And the Hubble parameter \eqref{M1} can be rewritten in terms of the scalar curvature $\R$,
\begin{equation}
H=H_0-H_1e^{-\beta t}=H_0\mp\sqrt{\tilde{R}-AH_0^2}.
\label{M5}
\end{equation}
In this case the first Friedmann equation \eqref{1.18} yields,
\begin{equation}
12\mu\beta (AH_0^2-\tilde{R}) (H_0\mp\sqrt{\tilde{R}-AH_0^2})F^{''}-BF^{'}+F-\kappa^2\rho_m=0\ ,
\label{M6}
\end{equation}
where $B=6[(1-3\lambda+3\mu)H^2+\mu\dot{H}]$. Then, by setting  $\beta=\frac{2H_0(1-3\lambda+3\mu)}{\mu}$, we obtain,
\begin{equation}
B=6(1-3\lambda+3\mu)(1-A)H_0^2+6(1-3\lambda+3\mu)\tilde{R}\ .
\label{M7}
\end{equation}
Eq.~\eqref{M6}  is still a very difficult expression, so that the search of  exact solutions for $F(\R)$ is a difficult task. 
Nevertheless, we can reconstruct some particular exact actions by considering special matter fluids. Let us consider the matter energy density,
\begin{equation}
\rho_m=\kappa^{-2}[12\mu\beta (AH_0^2-\tilde{R}) (H_0\mp\sqrt{\tilde{R}-AH_0^2})F^{''}-Ca^{-3}]\ .
\label{M8}
\end{equation}
Then the FLRW equation \eqref{M6} admits the following particular solution
\begin{equation}
F(\tilde R)=C_1[6(1-3\lambda+3\mu)\tilde{R}+6(1-3\lambda+3\mu)(1-A)H_0^2]^{\frac{1}{6(1-3\lambda+3\mu)}}.
\end{equation}
In a similar way, other particular solutions of the Friedmann equations can be reconstructed. Hence, we have shown here that ekpyrotic 
universes can be well described in the frame of Ho\v{r}ava-Lifshitz gravity.

\section{Discussions}
In the present paper, we have analyzed some particular cosmological solutions in the context of Ho\v{r}ava-Lifshitz gravity, where basically 
some generalizations of the original action \cite{Horava}, similar to standard $F(R)$ gravity, have been studied. It is well known that for a 
particular Hubble parameter, the corresponding action can be reconstructed in the framework of $F(\R)$ Ho\v{r}ava-Lifshitz gravity (see 
Ref~\cite{Elizalde:2010ep}), where the presence of the set of parameters $\{\lambda,\mu\}$, consequence of the restriction of the symmetries of 
the theory, can vary along the cosmological evolution, since their value depends on the energy scale of a particular system. Hence, the 
presence of this set of parameters will fluctuate along the universe evolution, affecting the corresponding cosmic solution. By assuming that 
GR should be  recovered when $\R\sim H^2_0<<m_{pl}^4\sim10^{74}GeV^4$, the parameters  $\lambda=\mu\sim1$ during the radiation/matter dominated 
epoch and the current accelerating era, while it becomes large when $\R\propto m_{pl}^4$, where the quantum effects should become important. In 
this sense, the effects of Ho\v{r}ava-Lifshitz gravity, and specifically the extra scalar mode, may become important when the universe reaches 
stages as the Little Rip, or other phases from a typical ekpyrotic universe \\ 

Hence, in the particular solutions studied here, the ekpyrotic scenario becomes an important focus for analyzing Ho\v{r}ava-Lifshitz gravity, 
as the universe owns a periodic behavior, crossing different stages, where the quantum nature of the theory may be relevant.  Moreover, we have 
shown that particular actions which lead to a cyclic nature of the Hubble parameter can be reconstructed.  Several techniques have been used 
for the reconstruction procedure. By using an auxiliary scalar field, coming from the $F(\R)$ sector, we have shown that cosmological solutions 
can be easily obtained. In addition, we have studied the shape of the action along each phase of a typical ekpyrotic universe, where the 
corresponding actions have been obtained. It is straightforward to show that such actions lead to standard $F(R)$ gravity when $\lambda=\mu=1$, 
and can be identified with some particular viable theories \cite{Nojiri:2009kx}.  Then, we can conclude that this class of actions can 
perfectly describe the entire universe evolution by means of an ekpyrotic model. Moreover, we have suggested the compatibility between an 
ekpyrotic universe and the presence of a Little Rip, a non singular point that may lead to the break of some bounded systems, where the effects 
of Ho\v{r}ava-Lifshitz gravity turn out important, and $\lambda\neq1,\mu\neq 1$. Future singularities can not be compatible with a cyclic 
universe unless a cure for the singularity is considered \cite{LopezRevelles:2011uc}. A next step should be to probe the possibility to reproduce cyclic cosmologies within the frame of  so-called viable $F(\tilde{R})$ gravities (see for instance, Ref.~\cite{Hu:2007nk}). While the violation of Newtonian law can be avoided in $F(\tilde{R})$ Ho\v{ra}va-Lifshitz  gravity (see \cite{Elizalde:2010ep}), the presence of instabilities and other features should be studied in more detail.\\

On the other hand, in order to have a complete picture of the universe evolution, one should specify how reheating occurs. Nevertheless, this is beyond of the scope of this work, but an interesting proposal for a reheating mechanism in the frame of UV complete theory is pointed out in \cite{Germani:2009yt}.\\

Therefore, in an ekpyrotic universe, the main implications of $F(\R)$ Ho\v{r}ava-Lifshitz gravity would come during those phases when the full 
diffeomorphisms are broken, basically during the early and ending phases, that may affect other {\it classical} eras, specially by the 
perturbations, which should be an important point to be studied in the future, where the effects may be distinguishable from other models.

\begin{acknowledgments}
We would like to thank the referee of a previous version for comments and criticisms that led to its improvement. AJLR acknowledges a JAE fellowship from CSIC. DSG acknowledges support from  a postdoctoral contract from the University of Basque Country.    
\end{acknowledgments}


\begin{thebibliography}{}

\bibitem{ekpyrotic}
J.~Khoury, B.~A.~Ovrut, P.~J.~Steinhardt and N.~Turok,
Phys.\ Rev.\ D {\bf 64}, 123522 (2001)
[arXiv:hep-th/0103239]; \\
J.~Khoury, B.~A.~Ovrut, P.~J.~Steinhardt and N.~Turok,
Phys.\ Rev.\ D {\bf 66}, 046005 (2002)
[arXiv:hep-th/0109050]; \\
P.~J.~Steinhardt and N.~Turok,
Science {\bf 312}, 1180 (2006)
[arXiv:astro-ph/0605173].

\bibitem{Elizalde:2008yf} 
  E.~Elizalde, S.~Nojiri, S.~D.~Odintsov, D.~S\'aez-G\'omez and V.~Faraoni,
  Phys.\ Rev.\ D {\bf 77}, 106005 (2008)
  [arXiv:0803.1311 [hep-th]].

 \bibitem{Elizalde:2010xq} 
  E.~Elizalde and A.~J.~Lopez-Revelles,
  Phys.\ Rev.\ D {\bf 82}, 063504 (2010)
  [arXiv:1004.5021 [hep-th]];

M.~Jamil, N.~A.~Myrzakulov, K.~K.~Yerzhanov, D.~Momeni and R.~Myrzakulov, 
arXiv:1201.4360 [physics.gen-ph];

R.~Myrzakulov, 
  \bibitem{Clifton:2011jh} 
  T.~Clifton, P.~G.~Ferreira, A.~Padilla and C.~Skordis,
  Phys.\ Rept.\  {\bf 513}, 1 (2012)
  [arXiv:1106.2476 [astro-ph.CO]].
  
\bibitem{review}
S.~Nojiri and S.~D.~Odintsov,
  eConf {\bf C0602061}, 06 (2006)
  [Int.\ J.\ Geom.\ Meth.\ Mod.\ Phys.\  {\bf 4}, 115 (2007)];
  Phys.\ Rept.\  {\bf 505}, 59 (2011)
  [arXiv:1011.0544 [gr-qc]].
  
S.~Capozziello and M.~Francaviglia,
  Gen.\ Rel.\ Grav.\  {\bf 40}, 357 (2008)
 [arXiv:0706.1146 [astro-ph]];
  
T.~P.~Sotiriou and V.~Faraoni, Rev. Mod. Phys. \textbf{82} 451 (2010);
  
S.  Capozziello, M.  De Laurentis, Phys.\ Rept.\ \textbf{ 509} 167 (2011) [arXiv:1108.6266 [gr-qc]];

R.~Myrzakulov, 
arXiv:1008.4486 [physics.gen-ph].
\bibitem{Nojiri:2006gh} 
  S.~Nojiri and S.~D.~Odintsov,
  Phys.\ Rev.\ D {\bf 74}, 086005 (2006)
  [hep-th/0608008].

\bibitem{SaezGomez:2008uj} 
  D.~S\'aez-G\'omez,
  Gen.\ Rel.\ Grav.\  {\bf 41}, 1527 (2009)
  [arXiv:0809.1311 [hep-th]].
  
\bibitem{Horava}
   P.~Ho\v{r}ava,
   Phys.\ Rev.\  D {\bf 79}, 084008 (2009)
   [arXiv:0901.3775 [hep-th]].

\bibitem{Charmousis:2009tc} 
  C.~Charmousis, G.~Niz, A.~Padilla and P.~M.~Saffin,
  JHEP {\bf 0908}, 070 (2009)
  [arXiv:0905.2579 [hep-th]].

\bibitem{Blas:2009yd} 
  D.~Blas, O.~Pujolas and S.~Sibiryakov,
  JHEP {\bf 0910}, 029 (2009)
  [arXiv:0906.3046 [hep-th]].


  \bibitem{Blas:2009qj}
  D.~Blas, O.~Pujolas and S.~Sibiryakov,
  Phys.\ Rev.\ Lett.\  {\bf 104}, 181302 (2010)
  [arXiv:0909.3525 [hep-th]].

\bibitem{Horava3}
   P.~Ho\v{r}ava, C.~M.~Melby-Thompson
  Phys.\ Rev.\ D {\bf 82}, 064027 (2010)
   [arXiv:1007.2410 [hep-th]].
   
   \bibitem{Kluson:2010za}
  J.~Kluson, S.~Nojiri, S.~D.~Odintsov and D.~S\'aez-G\'omez,
  Eur.\ Phys.\ J.\ C {\bf 71}  1690 (2011)
  [arXiv:1012.0473 [hep-th]].
   
   \bibitem{cosm}
   T.~Takahashi and J.~Soda,
   Phys.\ Rev.\ Lett.\  {\bf 102}, 231301 (2009)
   [arXiv:0904.0554 [hep-th]]; \\
   E.~Kiritsis and G.~Kofinas,
   Nucl.\ Phys.\  B {\bf 821}, 467 (2009)
   [arXiv:0904.1334 [hep-th]]; \\
   R.~Brandenberger,
   Phys.\ Rev.\  D {\bf 80}, 043516 (2009)
   [arXiv:0904.2835 [hep-th]]; \\
   S.~Mukohyama, K.~Nakayama, F.~Takahashi and S.~Yokoyama,
   Phys.\ Lett.\  B {\bf 679}, 6 (2009)
   [arXiv:0905.0055 [hep-th]]; \\
   T.~P.~Sotiriou, M.~Visser and S.~Weinfurtner,
   JHEP {\bf 0910}, 033 (2009)
   [arXiv:0905.2798 [hep-th]]; \\
   E.~N.~Saridakis,
   Eur.\ Phys.\ J.\  C {\bf 67}, 229 (2010)
   [arXiv:0905.3532 [hep-th]]; \\
   M.~Minamitsuji,
   Phys.\ Lett.\  B {\bf 684}, 194 (2010)
   [arXiv:0905.3892 [astro-ph.CO]]; \\
   G.~Calcagni,
   Phys.\ Rev.\  D {\bf 81}, 044006 (2010)
   [arXiv:0905.3740 [hep-th]]; \\
   A.~Wang and Y.~Wu,
   JCAP {\bf 0907}, 012 (2009)
   [arXiv:0905.4117 [hep-th]]; \\
   M.~i.~Park,
   JHEP {\bf 0909}, 123 (2009)
   [arXiv:0905.4480 [hep-th]]; \\
   S.~Nojiri and S.~D.~Odintsov,
   Phys.\ Rev.\  D {\bf 81}, 043001 (2010)
   [arXiv:0905.4213 [hep-th]]; \\
   M.~Jamil, E.~N.~Saridakis and M.~R.~Setare,
   Phys.\ Lett.\  B {\bf 679}, 172 (2009)
   [arXiv:0906.2847 [hep-th]]; \\
   M.~i.~Park,
   JCAP {\bf 1001}, 001 (2010)
   [arXiv:0906.4275 [hep-th]]; \\
   C.~Bogdanos and E.~N.~Saridakis,
   Class.\ Quant.\ Grav.\  {\bf 27}, 075005 (2010)
   [arXiv:0907.1636 [hep-th]]; \\
   C.~G.~Boehmer and F.~S.~N.~Lobo,
   arXiv:0909.3986 [gr-qc]; \\
   I.~Bakas, F.~Bourliot, D.~Lust and M.~Petropoulos,
   Class.\ Quant.\ Grav.\  {\bf 27}, 045013 (2010)
   [arXiv:0911.2665 [hep-th]]; \\
   G.~Calcagni,
   JHEP {\bf 0909}, 112 (2009)
   [arXiv:0904.0829 [hep-th]]; \\
   S.~Carloni, E.~Elizalde and P.~J.~Silva,
   Class.\ Quant.\ Grav.\  {\bf 27}, 045004 (2010)
   [arXiv:0909.2219 [hep-th]]; \\
   X.~Gao, Y.~Wang, W.~Xue and R.~Brandenberger,
   JCAP {\bf 1002}, 020 (2010)
   [arXiv:0911.3196 [hep-th]]; \\
   Y.~S.~Myung, Y.~W.~Kim, W.~S.~Son and Y.~J.~Park,
   arXiv:0911.2525 [gr-qc]; \\
   E.~J.~Son and W.~Kim,
   arXiv:1003.3055 [hep-th]; \\
   A.~Wang,
   arXiv:1003.5152 [hep-th]; \\
   A.~Ali, S.~Dutta, E.~N.~Saridakis and A.~A.~Sen,
   arXiv:1004.2474 [astro-ph.CO];\\
 S.~Mukohyama, arXiv:1007.5199 [hep-th];\\
E.~N.~Saridakis,
  Int.\ J.\ Mod.\ Phys.\ D {\bf 20}, 1485 (2011)
  [arXiv:1101.0300 [astro-ph.CO]];\\
  G.~Nugmanova, S.~.R.~Myrzakul, O.~Razina, K.~Esmakhanova, N.~Serikbayev and R.~Myrzakulov,
  arXiv:1104.5374 [physics.gen-ph];\\
  Y.~S.~Piao, 
Phys.\ Lett.\ B {\bf 681}, 1 (2009), [arXiv:0904.4117 [hep-th]].
\bibitem{FRhorava}
   M.~Chaichian, S.~Nojiri, S.~D.~Odintsov, M.~Oksanen and A.~Tureanu, Class.\ Quantum\ Grav.\ {\bf 27}, 185021 (2010)
   [arXiv:1001.4102 [hep-th]]; 
   S.~Carloni, M.~Chaichian, S.~Nojiri, S.~D.~Odintsov, M.~Oksanen and
A.~Tureanu, Phys.\ Rev.\ D {\bf 82}, 065020 (2010)
   [arXiv:1003.3925 [hep-th]]; 

   J.~Kluson,
   Phys.\ Rev.\  D {\bf 81}, 064028 (2010)
   [arXiv:0910.5852 [hep-th]]; 

   J.~Kluson,
   arXiv:1002.4859 [hep-th];

  D.~S\'aez-G\'omez,
  Phys.\ Rev.\ D {\bf 83}  064040 (2011)
  [arXiv:1011.2090 [hep-th]].

  \bibitem{Elizalde:2010ep}
  E.~Elizalde, S.~Nojiri, S.~D.~Odintsov and D.~S\'aez-G\'omez,
  Eur.\ Phys.\ J.\ C {\bf 70}  351 (2010)
  [arXiv:1006.3387 [hep-th]].

\bibitem{Nojiri:2011kd} 
  S.~Nojiri, S.~D.~Odintsov and D.~S\'aez-G\'omez,
  arXiv:1108.0767 [hep-th].
   
 \bibitem{LittleRip}
  P.~H.~Frampton, K.~J.~Ludwick and R.~J.~Scherrer,
  Phys.\ Rev.\ D {\bf 84}, 063003 (2011)
  [arXiv:1106.4996 [astro-ph.CO]];
P.~H.~Frampton, K.~J.~Ludwick, S.~Nojiri, S.~D.~Odintsov and R.~J.~Scherrer,
arXiv:1108.0067 [hep-th];
P.~H.~Frampton, K.~J.~Ludwick and R.~J.~Scherrer,
  arXiv:1112.2964 [astro-ph.CO].
  
    \bibitem{LopezRevelles:2011uc} 
  A.~J.~Lopez-Revelles and E.~Elizalde,
  arXiv:1104.1123 [hep-th].
  
    \bibitem{Turok:2004gb} 
  N.~Turok, M.~Perry and P.~J.~Steinhardt,
  Phys.\ Rev.\ D {\bf 70}, 106004 (2004)
  [Erratum-ibid.\ D {\bf 71}, 029901 (2005)]
  [hep-th/0408083];
  
  G.~Niz and N.~Turok,
  Phys.\ Rev.\ D {\bf 75}, 026001 (2007)
  [hep-th/0601007];
  
  J.~L.~Lehners, P.~McFadden and N.~Turok,
  Phys.\ Rev.\ D {\bf 75}, 103510 (2007)
  [hep-th/0611259];
  
  B.~Craps, T.~Hertog and N.~Turok,
  arXiv:0712.4180 [hep-th];
  
  E.~J.~Copeland, G.~Niz and N.~Turok,
  Phys.\ Rev.\ D {\bf 81}, 126006 (2010)
  [arXiv:1001.5291 [hep-th]].
  
  \bibitem{Buchbinder:2007ad} 
  E.~I.~Buchbinder, J.~Khoury and B.~A.~Ovrut,
  Phys.\ Rev.\ D {\bf 76}, 123503 (2007)
  [hep-th/0702154].
  
\bibitem{ADM}
R. L. Arnowitt, S. Deser and C. W. Misner, arxiv:gr-qc/0405109;
C.~Gao,
    Phys.\ Lett.\  B {\bf 684}, 85 (2010)    [arXiv:0905.0310 [astro-ph.CO]].

\bibitem{Mukohyama:2009mz} 
  S.~Mukohyama,
  Phys.\ Rev.\ D {\bf 80}, 064005 (2009)
  [arXiv:0905.3563 [hep-th]].
  
  \bibitem{Nojiri:2009kx} 
  S.~Nojiri, S.~D.~Odintsov and D.~S\'aez-G\'omez,
  Phys.\ Lett.\ B {\bf 681}, 74 (2009)
  [arXiv:0908.1269 [hep-th]].

  \bibitem{Khoury:2004xi} 
  J.~Khoury,
  astro-ph/0401579.
  
\bibitem{Linde:2002ws} 
  A.~D.~Linde,
  hep-th/0205259.

\bibitem{Spergel:2006hy}
D.~N.~Spergel {\it et al.} [WMAP Collaboration],
Astrophys.\ J.\ Suppl.\ {\bf 170}, 377 (2007)
[arXiv:astro-ph/0603449].

\bibitem{Hu:2007nk} 
  W.~Hu and I.~Sawicki,
  Phys.\ Rev.\ D {\bf 76}, 064004 (2007)
  [arXiv:0705.1158 [astro-ph]].

\bibitem{Germani:2009yt} 
  C.~Germani, A.~Kehagias and K.~Sfetsos,
  JHEP {\bf 0909}, 060 (2009)
  [arXiv:0906.1201 [hep-th]].
  
\end{thebibliography}
\end{document}